\documentclass[aps,physrev,twocolumn]{revtex4-2}

\usepackage{graphicx}
\usepackage{dcolumn}
\usepackage{bm}

\usepackage{braket}
\usepackage{amsmath}
\usepackage{float}
\usepackage{soul}

\usepackage[usenames,dvipsnames]{color}
\preprint{APS/123-QED}

\begin{document}

\title{\textbf{Phase modulation detection of a strontium atom interferometer gyroscope} }

\author{Luke A. Kraft}

\author{Samuel A. Meek}
 
\author{Nathan Marliere}

\author{Akbar Jahangiri Jozani}\thanks{Present address: Institute for Quantum Computing (IQC) and Department of Physics and Astronomy, University of Waterloo, Waterloo, Ontario, N2L 3G1, Canada}

\author{Grant W. Biedermann}
\email{Contact author: biedermann@ou.edu}
\affiliation{
  Homer L. Dodge Department of Physics and Astronomy, University of Oklahoma, Norman, Oklahoma, USA
}
\affiliation{Center for Quantum Research and Technology (CQRT), University of Oklahoma, Norman, Oklahoma, USA}

\date{\today}

\begin{abstract}
We demonstrate a strontium thermal beam atom interferometer gyroscope (AIG) on a rotation table using the $^{1\!}S_0$-$ ^{3\!}P_1$ intercombination line, and measure large rotation rates exceeding 6 rad/s. 
Our demonstration relies upon a transit-time-resonant (TTR), phase modulation technique for detecting the AIG  phase which rejects signal background and variations in fringe amplitude.  
\end{abstract}

\maketitle

\section{\label{Intro}Introduction}

The exceptional performance of light-pulse atom interferometers~\cite{kasevich91, peters01, gustavson00, mcguirk99} (LPAI) has motivated the development of LPAI for fundamental investigations~\cite{tino21rev} and applications worldwide~\cite{bongs19}.  
Campaigns at the frontiers of science include tests of the Einstein Equivalence Principle~\cite{asenbaum20,  yuan23, tino20}, measurements of the gravitational constant~\cite{jain21, fixler07} and the fine structure constant~\cite{parker18, morel20}, the development of gravitational wave detectors~\cite{abe21, canuel18}, the search for dark matter~\cite{du22, tino21rev} and dark energy~\cite{jaffe17, sabulsky19}, and LPAI with entangled matter~\cite{greve22, cassens24, huang24, brif20, hosten16}. 
In applications such as inertial sensing, a continual challenge is the realization of compact, high-performance, and robust rotation sensors~\cite{jekeli2023}.
Atom interferometer gyroscopes (AIG), using both cold~\cite{muller09, DArmagnac_de_Castanet24, Rakholia2014, stockton11, dutta16, tackmann14} and thermal beam approaches~\cite{durfee06, gustavson97, riehle91}, have inspired further research and development aimed at fielding this technology~\cite{Narducci22, black23, Meng24}.
Of these, thermal beam AIGs are intrinsically less complex and more robust in dynamic environments, making them attractive for field applications.

Thermal beam AIG approaches typically employ a two-photon stimulated Raman transition~\cite{Berman_1997} in alkali atoms~\cite{gustavson00,Narducci22, black23}. 
These experiments shoulder inherent technical complexity such as fast microwave modulation of the optical field, limited pulse efficiency and state preparation requirements~\cite{gustavson00}, obfuscating the intrinsic value of this approach. 
In adjacent work, gravity measurements with optical clock atom interferometers using laser-cooled alkaline earth atoms~\cite{katori99, miyake19, ovsiannikov16, rudolph20, santra05, wilkason22, xu03, ye06}  show key advantages of a direct optical transition to metastable states.  
We note that in strontium-88, the single ground state and the optical clock $^{1\!}S_0$-$ ^{3\!}P_1$ intercombination line can greatly simplify the laser system approach for a thermal beam AIG and reduce optical power requirements~\cite{rudolph20}.  Here, we demonstrate a strontium thermal beam atom interferometer gyroscope that leverages these distinct advantages in a minimal setup.  
Concurrently, we introduce a transit-time-resonant (TTR) phase modulation detection technique that significantly extends the AIG dynamic range and rejects certain systematic and statistical noise sources, allowing us to accurately measure rotation rates nearing 1 revolution per second.  

\begin{figure}
    \includegraphics[width=\linewidth, alt={Conceptual diagram of TTR phase modulation detection.  Atoms traversing the interferometer region receive a phase imprint, $\phi_i$, from the 689 nm laser field at each beam which is modulated at a frequency resonant with the inverse of the atom transit time, $2T = 2L/v_z$.  Depending on the time at which atoms enter the interferometer, the LPAI phase shift  varies between $\Delta\phi_{\rm{mod}}= \pm 4m$ as shown.  This modulated phase shift is detected in the fluorescence signal during readout and demodulated to acquire the inertial phase shift $\Delta\phi^{(I)}$ of the AIG.}]{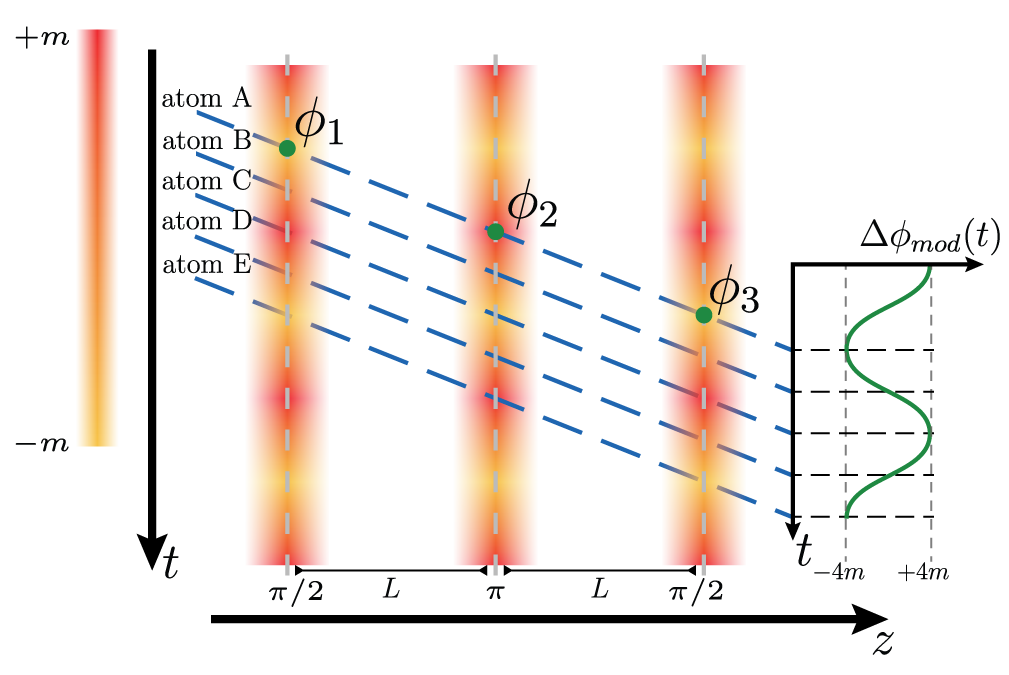}
    \caption{Conceptual diagram of TTR phase modulation detection.  Atoms traversing the interferometer region receive a phase imprint, $\phi_i$, from the 689 nm laser field at each beam which is modulated at a frequency resonant with the inverse of the atom transit time, $2T = 2L/v_z$.  Depending on the time at which atoms enter the interferometer, the LPAI phase shift  varies between $\Delta\phi_{\rm{mod}}= \pm 4m$ as shown.  This modulated phase shift is detected in the fluorescence signal during readout and demodulated to acquire the inertial phase shift $\Delta\phi^{(I)}$ of the AIG. 
}
    \label{fig:phaseDiagram}
\end{figure}

Similar to the first thermal beam matter-wave interferometer demonstration~\cite{riehle91}, our AIG employs a simple three-step LPAI process without explicit state preparation. 
Here, the strontium-88 atoms emerge collimated from an effusive oven in the pure quantum ground state of $^{1\!}S_0$,  transit to the LPAI region where an optical phase is imprinted (see Fig.~\ref{fig:phaseDiagram}), and then to the detection region for state readout.  
The LPAI region consists of three consecutive, evenly spaced, continuous wave beams of resonant light near 689~nm tuned to $^{1\!}S_0$-$ ^{3\!}P_1$, with the beam waists and intensities set to create a $\pi/2-\pi-\pi/2$ light pulse atom interferometer sequence~\cite{Berman_1997}. 

To first order~\cite{bongs06}, the resulting phase shift is given by $\Delta\phi = \phi_1 - 2\phi_2 + \phi_3$ where $\phi_i = \vec{k}\cdot\vec{x}_i + \phi_{m}(t_i)$ is the optical phase discretely sampled by the atoms at each of the three LPAI beams.  Here $\vec{k}$ is the wavevector of the 689~nm laser, $\vec{x}_i$ is the position of the atom in the $i^{th}$ pulse, and $\phi_{m}(t_i)$ is an imposed modulation of the optical field also sampled in the $i^{th}$ pulse.  The second-order central finite difference form of $\Delta\phi$ makes the interferometer sensitive to inertial dynamics such as acceleration and rotation which cause a curvature of the atom trajectory. In our setup, we also take advantage of this response to create our TTR phase modulation transfer effect to detect $\Delta\phi$.

TTR phase modulation detection in an LPAI leverages resonant enhancement of the modulation response for a specific atom velocity, $v$.
Consider an imposed sinusoidal phase modulation of the optical field, $\phi_{m}(t) = m \cos(\omega_{m} t)$ where $m$ is the modulation depth.  For a distance $L$ between the LPAI beams, setting $\omega_{m}=\pi\,v/L$ maximizes the response as follows.
Depending on the time at which such an atom enters the LPAI zone, the AIG response varies between $\Delta\phi = \pm 4m$ (see Fig.~\ref{fig:phaseDiagram}).  
This modulation of $\Delta\phi$ dithers the sinusoidal response of the LPAI and hence the detection signal.  
If the AIG phase caused by inertial effects is near $\Delta\phi^{(I)} = q\pi$, $\left(\Delta\phi^{(I)}  = (q+1/2)\pi\right)$, where $q$ is an integer, a tone at $2\omega_{m}$, ($\omega_{m}$), is produced, respectively. The ratio of the signals at these two frequencies normalizes the AIG fringe in real time and immunizes the system against dynamic changes in signal size and fringe contrast~\cite{Bohm1983, Kawasaki25} while simultaneously suppressing the response of non-resonant atoms.

Multiple techniques have been invented to measure the phase of an LPAI~\cite{kasevich91}.  
These range from simple fluorescence measurements~\cite{riehle91}, to sophisticated methods involving modulation transfer~\cite{McGuirk01, Hardman:16, gustavson00, black23} and normalized detection of spatially separated output ports~\cite{Biedermann:09, McGuirk01}.  
Normalization is crucial for achieving the best signal-to-noise ratio (SNR)~\cite{McGuirk01, Biedermann:09, Santarelli99} and is a common element for high-performance experiments, which often leverage low background signals for the approach. 
In a simple thermal beam AIG, many of the atoms emerging from the oven fall outside the ideal velocity conditions for the LPAI, and a large signal background exists. A common approach to mute the impact of background signals and their deleterious drifts is via phase sweep of one of the three LPAI beams~\cite{black23, gustavson00}.  
This modulates the fluorescence signal at a multiple of the sweep frequency and transfers the interferometer phase information, $\Delta\phi$, into a quiet frequency band.  This approach also scans the entire fringe and is thus self-normalizing.  However, modulating a single beam can also present a practical challenge, and often introduces additional phase noise and thermally-induced drifts from active phase control elements~\cite{gustavson00phd}.  
To circumvent these challenges, we implement phase modulation in common to all three LPAI beams allowing common-mode rejection of the added noise.   This approach also resonantly optimizes the signal from a specific velocity class thereby introducing a measure of control over the AIG response. 
In what follows, we describe this phase modulation detection technique in more detail in Sec.~\ref{LPAI}, present the features of our apparatus in Sec.~\ref{App}, and report the results of our experiment in Sec.~\ref{Results}. 

\section{\label{Theory}Theory of transit-time-resonant phase modulation detection in light pulse atom interferometers}
\label{LPAI}

Atom interferometers invert the role of matter and light, measuring phase differences between two propagating atomic wave packets traversing an interferometric loop defined by light pulses. 
In our thermal beam AIG, the light pulses are realized by atoms transiting a region of spatially separated, resonant laser fields.
The resonant interactions coherently separate, redirect, and recombine the strontium wave packets and sequentially imprint the spatially dependent phase of the light field, $\phi_i^{(I)} = \vec{k}\cdot\vec{x}_i$, onto the difference phase of the superposition state.

The resulting inertial phase response of the LPAI, $\Delta\phi^{(I)} = \phi_1^{(I)} - 2\phi_2^{(I)} + \phi_3^{(I)}$,  is sensitive to path curvature and responds to rotation according to  $\Delta\phi^{(I)}=-2\vec{k} \cdot (\vec{v} \times \vec{\Omega}) T^{2}$ where $T$ is the time between interactions and $\vec{v}$ is the velocity of the atom.  
For a physical laser field separation of $L$, we have $T=L/v_z$ where $v_z=\vec{v}\cdot\hat{z}$ (see Fig.~\ref{fig:setup}), and the AIG response can be written as $\Delta\phi^{(I)} = - 2 \vec{k} \cdot (\vec{v} \times \vec{\Omega}) (L/v_z)^{2}$. Here, we neglect the additive phase shift due to acceleration, $\Delta\phi_{acc}=\vec{k}\cdot\vec{a}\,(L/v_z)^2$, since it is small when compared to the rotational effect. 

We maximize the sensitivity of the AIG in our experiment by configuring $T$ to be comparable to $\tau$, the lifetime of $^3\!P_1$, the excited interferometer state.  Hence, it is important to include spontaneous decay in the analysis of the LPAI signal.
We detect the LPAI atoms with light-induced fluorescence at 461~nm on the $^{1\!}S_0$-$ ^{1\!}P_1$ transition providing a signal proportional to the probability in $^{1\!}S_0$.  Assuming a perfect $\pi/2 - \pi- \pi/2$ pulse sequence in the interferometer, the probability of the atom to be in the $^{1\!}S_0$ state upon reaching the detection laser is given by 
\begin{equation}
    P(t) = \frac{1}{2}e^{-T'/\tau}\cos{\left(\Delta\phi_{\rm{tot}}(t)\right)} + \Bigl(1-\frac{1}{2}e^{-(T'-T)/\tau}\Bigr)\rm{,}
    \label{eqn:pt}
\end{equation}
where $\Delta\phi_{\rm{tot}}(t) = \Delta\phi^{(I)}(t)+\Delta\phi_{\rm{mod}}(t)$ is the total phase response of the interferometer including both inertial and modulation effects, and $T'$ is the transit time from the midpoint of the interferometer to the detection laser. 
A derivation of this equation is included in the Appendix, but we note for clarity that this reduces to the familiar $P(t) = \frac{1}{2}\left(1+\cos{\left(\Delta\phi_{\rm{tot}}(t)\right)} \right)$ in the limit of $T' \ll \tau$.
Adding modulation to $\Delta\phi^{(I)}$ dynamically scans the AIG phase response creating fluorescence signal tones proportional to the slope and curvature of the LPAI fringe as previously noted.

If we modulate the 689 nm laser phase with $\phi_{\rm{mod}}(t) = m \cos(\omega_{m} t)$, $\Delta\phi_{\rm{tot}}$ will then respond as
\begin{equation}
\begin{aligned}
    \Delta\phi_{\rm{tot}}(t+T') = \,\,&\phi_1^{(I)} + m \cos(\omega_{m} (t - T)) \\
                       - 2 (&\phi_2^{(I)} + m \cos(\omega_{m} t)) \\
                          + &\phi_3^{(I)} + m \cos(\omega_{m} (t + T))\rm{.}
\end{aligned}
\end{equation}

Using trigonometric substitution and regrouping terms yields
\begin{equation}
    \Delta\phi_{\rm{tot}}(t+T') = \Delta\phi^{(I)} - m' \cos(\omega_{m}t)\rm{,}
    \label{eqn:dphitot}
\end{equation}
where
\begin{align}
    m' = 2m(1-\cos(\omega_{m} T)) = 4m \sin^2\!\left(\frac{\omega_m T}{2}\right) \label{eqn:mp}\rm{.}
\end{align}
Substituting equation (\ref{eqn:dphitot}) into equation (\ref{eqn:pt}) results in
\begin{equation}
\begin{aligned}
    P(t+T') = &\Bigl(1-\frac{1}{2}e^{-(T'-T)/\tau}\Bigr) \\ 
            + &\frac{1}{2}e^{-T'/\tau}\cos(\Delta\phi^{(I)})\cos(m' \cos(\omega_m t)) \\
            + &\frac{1}{2}e^{-T'/\tau}\sin(\Delta\phi^{(I)})\sin(m' \cos(\omega_m t))\rm{.}
\end{aligned}
\end{equation}
This can be expanded as a series in $\cos(n \omega_m t)$ using the Jacobi-Anger expansions
\begin{align}
    \cos(m' \cos(\omega_m t)) =&\\
        J_0(m') + 2\sum_{n=1}^{\infty}&(-1)^n J_{2n}(m') \cos(2n \omega_m t) \nonumber\\
    \sin(m' \cos(\omega_m t)) =&\\
        - 2\sum_{n=1}^{\infty}&(-1)^n J_{2n-1}(m') \cos((2n-1) \omega_m t) \nonumber.
\end{align}
If $P(t+T')$ is expressed as
\begin{equation}
    P(t+T') = \sum_{n=0}^{\infty} F_n \cos(n \omega_m t),
\end{equation}
then the first three terms are
\begin{alignat}{3}
    F_0 =&& \Bigl(1-\frac{1}{2}e^{-(T'-T)/\tau}\Bigr)& \nonumber\\
    &+&\frac{1}{2}e^{-T'/\tau}J_0(m')\cos&(\Delta\phi^{(I)})\\
    F_1 =& &e^{-T'/\tau}J_1(m') \sin&(\Delta\phi^{(I)})\label{eqn:F_1}\\
    F_2 =& &-e^{-T'/\tau}J_2(m') \cos&(\Delta\phi^{(I)}).\label{eqn:F_2}
\end{alignat}
Thus, $F_1$ and $F_2$ are proportional to $\sin(\Delta\phi^{(I)})$ and $\cos(\Delta\phi^{(I)})$, respectively.
Signals proportional to $F_1$ and $F_2$ can be recorded by demodulating the fluorescence signal at the respective frequency.
If appropriate scaling factors are applied to these two signals, such that they have the same peak amplitude, then $\Delta\phi^{(I)}$ can be measured directly.
Denoting the two scaled signals as $F_1'$ and $F_2'$, then $\Delta\phi^{(I)} = \operatorname{atan2}(F_1',F_2')$, where $\operatorname{atan2}$ is the two-argument arctangent function, equivalent to $\arctan(F_1'/F_2')$ when $F_2' > 0$. 
While the true $\operatorname{atan2}$ function has a branch cut at $\pm \pi$, we apply an unwrapping algorithm that adds multiples of $2 \pi$ to each point in the trace such that no two subsequent points differ by more than $\pm \pi$.  

\section{\label{App}Apparatus}

\begin{figure}
    \includegraphics[width=\linewidth, alt={System level overview of the experiment. The AIG sensor is mounted atop a precision turntable for rotation testing while the laser systems are located off table and fed to the experiment via fiber optics.  The 689~nm laser beamline is shown at the bottom of the figure. The 689~nm laser beamline is shown at the bottom of the figure.  This beam is divided and shaped into the $\pi/2$, $\pi$, and $\pi/2$ pulses that form the beam splitters of the LPAI.  The beamline of the 461~nm fluorescence detection laser is shown at the top of the figure.  Fluorescence from Sr atoms excited by this laser is collected on an avalanche photodiode (APD) module.  This signal is demodulated to calculate the time-dependent rotation rate $\Omega(t)$.}]{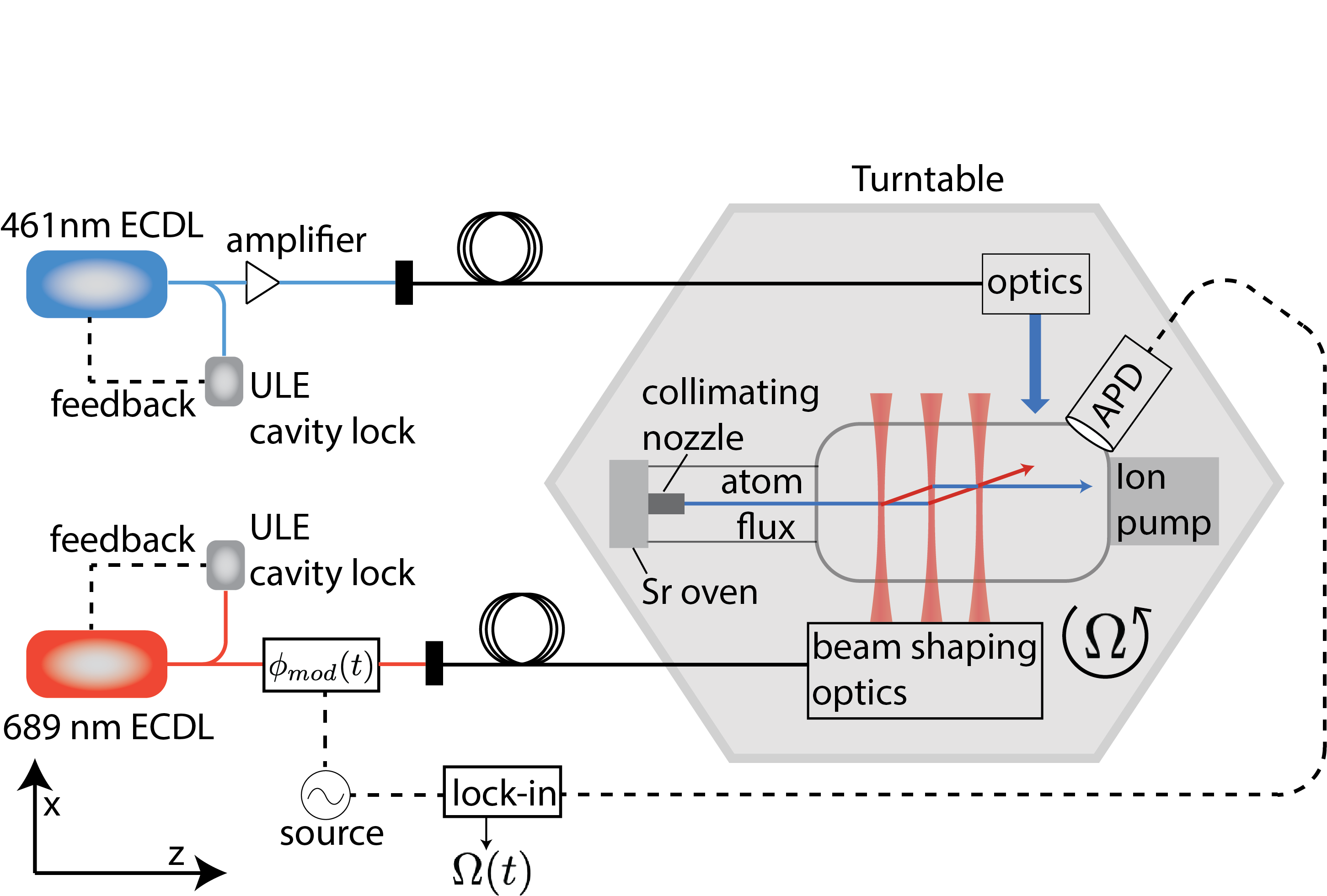}
    \caption{System level overview of the experiment. The AIG sensor is mounted atop a precision turntable for rotation testing while the laser systems are located off table and fed to the experiment via fiber optics.  The 689~nm laser beamline is shown at the bottom of the figure.  This beam is divided and shaped into the $\pi/2$, $\pi$, and $\pi/2$ pulses that form the beam splitters of the LPAI.  The beamline of the 461~nm fluorescence detection laser is shown at the top of the figure.  Fluorescence from Sr atoms excited by this laser is collected on an avalanche photodiode (APD) module.  This signal is demodulated to calculate the time-dependent rotation rate $\Omega(t)$.}
    \label{fig:setup}
\end{figure}

Our compact experiment consists of a thermal beam AIG mounted atop a rotating test platform and interrogated by two laser systems. 
As can be seen in Fig.~\ref{fig:setup}, the AIG begins with a heated strontium source that is collimated into a jet by a capillary array nozzle. 
This beam of atoms flies ballistically in vacuum to the interrogation zone where it interacts with  phase modulated 689~nm light, resonant with $^{1\!}S_0$-$ ^{3\!}P_1$, effecting the interferometer pulses. 
Immediately following the interferometer, a 461~nm beam, resonant with the $^{1\!}S_0$-$ ^{1\!}P_1$ transition, produces fluorescence monitored by an avalanche photodiode (APD). 
This fluorescence signal is  recorded together with a signal from a rotary encoder embedded within the turntable and processed in software to validate the measurement.

The vacuum chamber for the apparatus is 58~cm long and contains two primary regions: a Sr oven and the interferometer region. 
The oven produces a strontium vapor that is  collimated to a $\approx4\times10^{9}$~atom/s thermal beam via a micro-capillary array acting as the nozzle~\cite{senaratne15}.  
The nozzle is an equilateral triangular opening, 4~mm on a side, packed with 211~\textmu m  diameter by 8~mm long micro-capillary tubes. 
The capillaries restrict the divergence angle of the beam to 26~mrad and provide an estimated $\approx$~10~m/s FWHM transverse velocity profile.  
The interferometer region is constructed from an off-the-shelf, compact 2.75" ConFlat 2-way cross from Kimball Physics, with the two 2.75" windows oriented perpendicular to the thermal beam propagation for applying the interferometer and detection lasers. 
Immediately after the interferometer region is a vacuum tee connecting to a valve for pump out and an Omega 3~L/s ion pump for maintaining vacuum below a pressure of $10^{-6}$ Torr which is ample for the required coherence time of the interferometer.

Due to the low vapor pressure of alkaline-earth metals \cite{alcock84}, our atomic source uses a crucible maintained at a temperature of $420^\circ$C in order to provide sufficient flux.
The oven follows the design in \cite{senaratne15} and is filled with 99\% purity strontium granules.
The nozzle is kept at $\approx 440^\circ$C to prevent strontium deposition and clogging in the capillary array. 
This approach has proven to be effective in keeping the nozzle clear over approximately 2 years of continuous operation. 
This oven arrangement results in heating of the chamber components downstream from the nozzle. 
To prevent strontium migration through the vacuum system and coating of the viewports, a small 30.4~cfm fan and cowling have been installed to rapidly cool the reducing nipple connecting the oven to the rest of the vacuum chamber.  
To reduce thermal drift of the nearby optical components, an aluminum plate is mounted between the oven and the AIG optics to act as a heat reflector.

In the interrogation zone, the atoms pass through three $\pi$-polarized 689~nm beams that nominally provide $\pi/2-\pi-\pi/2$ LPAI pulse sequence on the $^{1\!}S_0$-$ ^{3\!}P_1(m=0)$ transition.
A bias magnetic field of $\approx$~110~G is applied along the laser polarization axis in the $\hat{y}$-direction (parallel to the rotation axis) using a permanent magnet on top of the chamber.  This defines the quantization axis and suppresses unwanted transitions to the $m=\pm1$ Zeeman sublevels.
To generate the three beams, a single 689~nm beam passes through a polarizing beam-splitting cube and divides into two paths. 
The transmitted light constitutes the $\pi$-beam, while the deflected component becomes the two $\pi/2$-beams after being further divided by a 50/50 non-polarizing beam-splitting cube. 
The three beams are then incident on three individual cylindrical lenses, one for each beam. 
These lenses are mounted with $\approx$~7~mm spacing so that, with each beam incident on the center of a lens, three foci with a measured spacing of $L=7.00(1)$~mm spacing are produced in the focal plane, which falls across the strontium thermal beam, and accommodates the $\tau = 21.3$~\textmu s lifetime of the $^{3\!}P_1$ excited state~\cite{sansonetti10}.
In the focal plane, each beam has a $1/e^2$ radius of $w_y = 1.942(52)$~mm in the vertical direction and $w_z = 42(6)$~\textmu m in the horizontal with measured powers of 1.43 mW, 7.14 mW, and 1.43 mW for the $\pi/2$, $\pi$, and $\pi/2$ beams respectively.
This entire optical layout is covered by a low-profile 3D-printed shroud to reduce the impact of air currents during rotation, and the base plate  is temperature stabilized using a heater attached to the underside.

Achieving an interferometric response requires precision alignment of the 689 nm LPAI beams. 
In the vertical direction, the parallelism must satisfy $| \bm{k}_1 - 2 \bm{k}_2 + \bm{k}_3| \ll 2 \pi/w_y\rm{,}$
where $\bm{k}_i$ is the wave vector of the $i^{th}$ beam, and $w_y$ is the radius of the vertical waist.  This gives a relative beam angle tolerance of $\alpha_{y} \ll \lambda / w_y \approx 400$~\textmu rad.
The horizontal alignment tolerance is bounded by the relative Doppler-shift between the interferometer beams such that each is resonant with the same velocity class.  This condition can be written as $|\bm{k}_i| \, v \, \alpha_z \ll   \,\Omega_r\rm{,}$

where $v$ the atom velocity and $\Omega_r$ is the Rabi frequency of the LPAI pulse on the $^{1\!}S_0$-$ ^{3\!}P_1$ transition.
Since the beam waist along the atomic beam axis is chosen to satisfy a $\pi$-pulse condition then $\Omega_r \approx v / w_z$.
Hence, the horizontal relative angular alignment tolerance is bounded by $\alpha_z \ll \lambda / w_z \approx 14$~mrad.  
In addition to these angular constraints, the spacing between the interferometer beams must be precisely controlled to close the loop of the interferometer to within the average coherence length of the wavepackets~\cite{biedermann17, Parazzoli12}. 
This results in a spacing tolerance of  $|L_{12}-L_{23}|\ll w_z$ where $L_{ij}$ is the distance between LPAI beam $i$ and $j$. 
We perform these alignments using paraxial lens theory~\cite{Pedrotti_Pedrotti_Pedrotti_2017} and a DataRay BladeCam2-XHR Beam Profiler mounted on a two-axis translation stage.  
With this method we achieve relative alignment uncertainties of $\sigma_{\alpha_y} = 1.2$~\textmu rad, $\sigma_{\alpha_z} = 2.7$~\textmu rad, and $|L_{12}-L_{23}| \leq 10$~\textmu m, satisfying the above constraints.

At the output of the interferometer, the population in the $^{1\!}S_0$ state is measured using fluorescence detection on the $^{1\!}S_0$-$ ^{1\!}P_1$ transition. 
A 461~nm beam is shaped using two cylindrical lenses and an edge mask, and directed through a viewport on the opposite side of the chamber, antiparallel to the interferometer beams.  

This light crosses the atomic beam immediately after the final beam of the interferometer, to minimize $T'-T$, the atom transit time to the detection zone, and hence the decay of $^{3\!}P_1$.
The previously mentioned magnetic field also sets the quantization axis for the excitation and fluorescent decay.
Since the laser polarization is aligned to this axis, fluorescence must be collected near the plane perpendicular to this axis for maximum signal.
In this apparatus, collection optics are placed just above the entering beam and direct fluorescence onto an avalanche photodiode module (Hamamatsu C12703-01) with a typical bandwidth of $f_{\rm{bw}}=100$~kHz.
The signal from this module is recorded using an FPGA-based data acquisition device (Red Pitaya STEMLab 125-14) running an adaption of the lock-in software package from Marcelo Luda~\cite{10.1063/1.5080345,Luda_Red_Pitaya_Lock-in_PID}.

The experiment requires two stabilized diode laser systems: one to generate the 689~nm beams of the LPAI and the other to generate 461~nm light for state-selective detection of atoms exiting the interferometer.
The 689~nm light is produced by a MOGLabs Cateye Laser system locked to a temperature stabilized, optical cavity constructed from ultra-low expansion (ULE) glass (Stable Laser Systems).
For this laser, we infer a linewidth of 1.5~kHz by analyzing the in-band error signal of the cavity lock \cite{Allan1990, xie17}.
The output of this laser is double-passed through an accousto-optic modulator (AOM) before transiting through a 7~m single-mode, polarization maintaining optical fiber to deliver 11 mW at 689 nm on platform and an estimated $\Omega_r\approx2\pi\times1$~MHz.
The AOM is used to phase modulate the 689 nm laser light by applying phase modulation to the radio frequency  drive.  
A MOGLabs Injection Lock Amplifier system provides 25~mW of 461~nm light on platform for driving the $^{1\!}S_0$-$ ^{1\!}P_1$ transition. 
The seed laser for this system is also locked to a temperature stabilized, ULE optical cavity (Stable Laser Systems) with a linewidth of $\approx$~1.6~MHz. 
The amplified light passes through a double-pass AOM to adjust the laser frequency onto resonance and is then fed into a 7~m single-mode fiber that outputs the light on the AIG platform. 

The entire experimental platform, including the vacuum chamber, beam shaping optics, and photodetector, is mounted on top of a precision turntable bearing (TPA Motion LLC FMB-100.180-N).
An incremental rotary encoder with 120,000 counts per revolution (Encoder Products Company 755A-19-S-30,000-R-PP-1-MF-J00-N) is attached between the fixed base and the rotatable experimental platform.
This provides a continuous measurement of the angular position of the platform, which we use to independently determine the angular velocity.
To read out the angular position, the two quadrature logic signals from the encoder are monitored and converted to a 16-bit incremental position by a microcontroller (Microchip ATmega328P on an Analog Devices Linduino development board).
This position value is converted to an analog signal using a 16-bit digital-analog converter (Analog Devices LTC2668-16 on a DC2025A-A evaluation board), which is recorded together with the fluorescence signal on the same data acquisition device.
The encoder and attached electronics are capable of tracking rotation rates of up to 3.3 revolutions per second (400,000 counts per second), limited by encoder's specified maximum frequency.
Practically, maximum rotation rates of approximately one revolution per second have been achieved, limited only by the method of manual rotation.
Although the bearing is capable of continuous rotation, the setup is currently restricted to peak-to-peak oscillations of roughly two revolutions due to fibers and cabling in the system.

\section{\label{Results}Results}

\begin{figure}
    \includegraphics[alt={Top: $F_2'$ signal vs. $F_1'$ signal (gray).  
    	An interferometer phase $\Delta\phi^{(I)} = \operatorname{atan2}(F_1',F_2')$ can be determined for each coordinate $(F_2',F_1')$ in the trace, as indicated in black.
    	Bottom: Time trace of the fluorescence signal demodulated at the modulation frequency ($F_1'$, dark red) and twice the modulation frequency ($F_2'$, light blue).}]{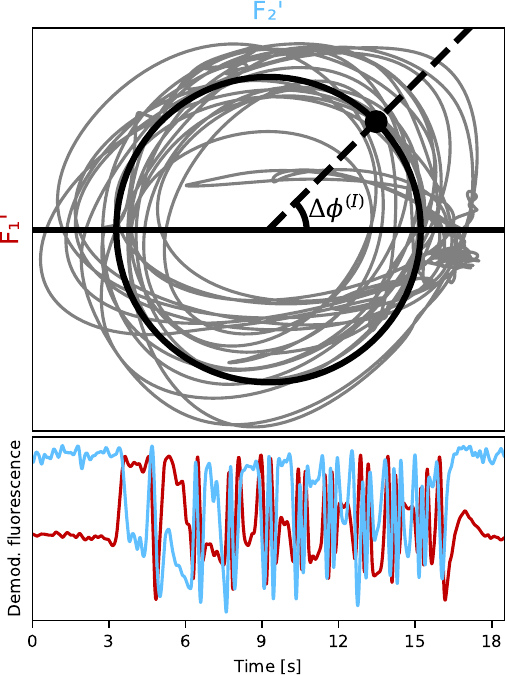}
    \caption{
        {\bf Top:} $F_2'$ signal vs. $F_1'$ signal (gray).  
        An interferometer phase $\Delta\phi^{(I)} = \operatorname{atan2}(F_1',F_2')$ can be determined for each coordinate $(F_2',F_1')$ in the trace, as indicated in black.
        {\bf Bottom:}~Time trace of the fluorescence signal demodulated at the modulation frequency ($F_1'$, dark red) and twice the modulation frequency ($F_2'$, light blue).  
    }
    \label{fig:f1f2}
\end{figure}

We characterize the AIG on our turntable using manual rotation and TTR phase modulation detection.   The LPAI laser is phase modulated at a frequency of $\omega_{\rm{mod}} = 2\pi\times(32768\,\rm{ns})^{-1} \approx 2\pi\times30.5$~kHz using a reference signal from a function generator.
The output of the function generator is fed to the phase modulation input of the RF synthesizer where it generates a phase modulation with a depth of $\pm0.3$~rad.  
This RF synthesizer drives the AOM for the interferometer laser.
Because the AOM operates in a double pass configuration, a total phase modulation depth of $m = 0.6$ is imprinted on the laser.

Analog measurement signals are recorded in two-channels with 14-bit resolution each at a sampling rate of $f_{\rm{samp}} = 8\times\omega_{\rm{mod}}/(2\pi) \approx 244$~kHz.
The fluorescence signal from the APD is recorded on channel 1, while the encoder signal is recorded on channel 2.
The software on the data acquisition device  is currently limited to simultaneous recording of two data channels, hence to record the modulation signal, a reference copy is multiplexed with the encoder signal going to channel 2 using a 100:1 resistive voltage divider.  These two signals are separated with post-processing. 
To calibrate the conversion factor between the recorded encoder signal and relative position of the platform, a digital pulse sequence from a function generator is applied to the inputs of the microcontroller board in place of the rotary encoder output, mimicking the platform rotating at a known, constant rate.

The two channels contain four signals of interest: fluorescence modulation at $\omega_{\rm{mod}}$ and $2\,\omega_{\rm{mod}}$ on channel 1, rotary encoder signals in the base band on channel 2, and the modulation reference signal at $\omega_{\rm{mod}}$ on channel 2.  
Each of these signals is extracted from the time trace using exactly the same procedure.
The time trace is first multiplied by $\exp(-i\,n\,\omega_{\rm{mod}} t)$, where $n = 0,1,2$ is the relevant harmonic, to shift the signal of interest to zero frequency after which a finite impulse response (FIR) low pass filter is applied to reject the other signals and unwanted images.
The $n-1$ taps of the FIR are given by
\begin{equation}
a_i = \biggl(1-a\cos\Bigl(\frac{2 \pi i}{n}\Bigr)+(a-1)\cos\Bigl(\frac{4 \pi i}{n}\Bigr)\biggr)/n
\end{equation}
where $i$ spans from 1 to $n-1$ and $a = (2\pi^2/9 - 1/3)$.
These coefficients are chosen so that $a_0 = a_n = 0$ (ensuring a $1/f^3$ roll off outside the pass band) and $a$ is chosen specifically so that the second derivative of the transfer function is zero at $f=0$, providing a maximally flat pass band.
Since the filter is symmetric ($a_i = a_{n-i}$), it also has a linear phase response.

For the current analysis, $n = 100001$, resulting a 3-dB bandwidth of $f_{3\,\rm{dB}} = 1.5475f_{\rm{samp}}/n = 3.78$~Hz.
The same filter is applied to each signal to ensure equal time delays.

The modulated fluorescence signals at $\omega_{\rm{mod}}$ and $2\,\omega_{\rm{mod}}$ are phase shifted by a factor of $\exp(i n \phi)$, where $\phi$ is the phase of the reference signal and $n = 1$ and $2$, respectively, for the two harmonics.
The signals are then shifted by an additional factor of $\exp(-i\,n\,\omega_{\rm{mod}} T')$ to account for the time delay from the central beam of the interferometer to the detection beam.
Best results were found by assuming a time delay of $T' = 20$~\textmu s, which corresponds to a $L' = 11.5$~mm distance of the detection laser from the central interferometer beam.
The real parts of the two signals are then divided by $J_n(m')$, where $m' = 0.6 \sin^2(\omega_{\rm{mod}} \,T/2) = 0.5$ based on an estimated $T$ of 12~\textmu s.
This produces the signals $F_1'$ and $F_2'$, which are proportional to $\sin(\Delta\phi^{(I)})$ and $\cos(\Delta\phi^{(I)})$, respectively.
Empirically, it was found that $F_2'$ needed to be multiplied by an additional factor of 2 to match its amplitude to the amplitude of $F_1'$; we attribute the need for this extra factor to low pass filtering effects from the non-zero width of the detection laser and the frequency rolloff of the APD detector.

The bottom part of Figure \ref{fig:f1f2} shows the $F_1'$ and $F_2'$ signals plotted as a function of time while manually applying an oscillation to the platform.  
The extrema of the signals are associated with the 2$\pi$ cyclic nature of the quadrature components of $F_1$ and $F_2$ in Eqs.~\ref{eqn:F_1}~\&~\ref{eqn:F_2}.  
Each signal, taken individually, suffers degradation in response fidelity near the peaks of the fringe.
In the top part of the figure, $F_2'$ is plotted versus $F_1'$ showing the signals tracing out an approximately circular pattern.
The inherent correlation of the two signals allows continuous access to the interferometer phase for all values of $\Delta\phi^{(I)}$ as schematically indicated in black.  Although the fringe amplitude changes significantly in this experiment due to phase broadening effects~\cite{Rakholia2014}, and $\Delta\phi^{(I)}$ spans more than $5\pi$ rad, the phase correlation between the two frequency components is preserved and effectively normalizes the readout in real time.

\begin{figure}
    \includegraphics[alt={Rotation rate $\Omega$ as determined by the AIG vs the rotation rate determined by a rotary encoder at the base of the platform showing accurate measurements at rates in excess of 6 rad/s while the fringe amplitude varies by a factor of 3.}]{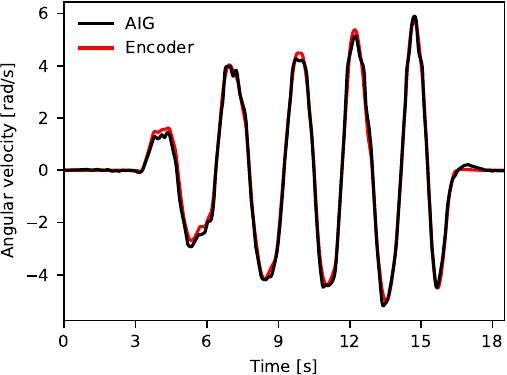}
    \caption{
      Rotation rate $\Omega$ as determined by the AIG vs the rotation rate determined by a rotary encoder at the base of the platform showing accurate measurements at rates in excess of 6 rad/s while the fringe amplitude varies by a factor of 3.   
    }
    \label{fig:encatan}
\end{figure}

Figure \ref{fig:encatan} shows the rotation rate determined using the AIG $\Omega_A$ compared to the rate determined using the rotary encoder $\Omega_E$.
The rate estimated from the AIG is calculated using 
\begin{equation}
\Omega_A = \operatorname{atan2}(F_1',F_2')\, \bar{v}_{\rm{eff}}/(2 k L^2)\rm{,}
\label{eqn:omega_a}
\end{equation}
with values of $\bar{v}_{\rm{eff}} = 580$~m/s for the effective mean longitudinal velocity of atoms participating in the interferometer, $k = 2 \pi \times 1450434\, \textrm{m}^{-1}$ for the angular wavenumber of the excitation beam, and $L = 7\, \textrm{mm}$ for the distance between excitation beams along the atomic beam axis.
A small offset is applied to $\Omega_A$ to correct for small residual interferometer phase at zero rotation rate.
The rate estimated by the encoder $\Omega_E$ is determined from a numerical derivative of the encoder position.
The plot shows excellent agreement between AIG and encoder rates even though the fringe amplitude is seen to vary by more than a factor of 3 in Fig.~\ref{fig:f1f2}.

While the value of $\bar{v}_{\rm{eff}}$ used above is determined empirically, it can also be estimated based on experimental parameters.
The on-axis flux of the atomic beam has a normalized distribution as a function of velocity that is given by
\begin{equation}
f(v)\,dv = \frac{v^3}{2 v_0^4} \exp\biggl(-\frac{v^2}{2 v_0^2}\biggr)\,dv\rm{,}
\label{eqn:fvdv}
\end{equation}
where $v_0 = \sqrt{k T / m}$.
The factor of $v^3$ accounts for the density to flux conversion and the spreading in the two transverse directions, each of which favor faster atoms proportional to $v$.
The mean velocity of this distribution can be calculated from 
\begin{equation}
\bar{v} = \int_0^{\infty} v f(v)\,dv = v_0 \sqrt{\frac{9 \pi}{8}}\rm{.}
\label{eqn:vbar}
\end{equation}
Using this mean velocity directly in equation \ref{eqn:omega_a}  overestimates the contribution of lower velocity atoms to the interferometer signal since these atoms are more likely to spontaneously decay from the $^3P_1$ state while passing through the interferometer. 
To account for this effect, the distribution in equation \ref{eqn:fvdv} is augmented with an extra exponential factor describing the survival probability of the excited state over the length of the interferometer.
\begin{equation}
f'(v) = f(v) \frac{\sqrt{\pi}}{G_{0, 3}^{3, 0}\left(\begin{matrix} - \\0, \frac{1}{2}, 2\end{matrix} \middle| {\frac{L'^2}{8{v_0}^2 \tau^2}} \right)} \exp\biggl(-\frac{L'}{v \tau}\biggr)
\end{equation}
Here, $L'$ is the estimated distance between the midpoint of the interferometer and the detection laser beam, $\tau$ is the lifetime of the $^3P_1$ state, and $G$ used in the normalization factor is the Meijer $G$-function.
Based on the $T = 440^\circ$C nozzle temperature, $v_0$ is 260~m/s, which corresponds to a mean velocity of $\bar{v} = 489$~m/s.
Including the decay effects with $L' = 11.5$~mm and $\tau = 21.3$~\textmu s \cite{sansonetti10} results in an estimated effective mean velocity which is somewhat higher, $\bar{v}_{\rm{eff,est}} = 560$~m/s.
Other factors, such as the optimum velocity for the desired $\pi/2-\pi-\pi/2$ pulse sequence or the transit time dependence of $m'$ in equation \ref{eqn:mp}, are not included in this calculation and can potentially further shift the effective mean velocity.

\section{\label{Conclusions}Outlook and Conclusion}

In conclusion, we have demonstrated an atom interferometer gyroscope using the $^{1\!}S_0$-$ ^{3\!}P_1$ intercombination line on a continuous thermal beam of strontium.  
Our demonstration leverages a transit-time-resonant phase modulation detection technique that allows readout for all values of $\Delta\phi^{(I)}$ across a range exceeding $5\pi$ radians and rejects certain spurious and systematic noise sources. 
This allows accurate readout of AIG phase information at rotation rates exceeding 6~rad/s and with accompanying significant changes in fringe amplitude.

As an outlook, it is informative to estimate the potential performance of an AIG based on this approach.  The fundamental limit to the angle random walk of an LPAI AIG is given by,
\begin{eqnarray}
    \sigma_{\Omega} ( t ) = \frac{v}{2 k L^2} \frac{1}{\sqrt{\mathcal{F} t}}
\end{eqnarray}
For an atom velocity of $v = 580~$m/s, $L = 7$~mm, and a flux of $\mathcal{F} \approx 10^{10}$~atoms/s, this system could yield navigation grade performance approaching $\sim$ \textmu rad/s/$\sqrt{\rm{Hz}}$.

Future research will aim to explore methods for increasing the sensitivity and improving the resolution of the AIG system we have developed. 
Possible methods include extended coherence times using the long-lived $^{3\!}P_0$ state~\cite{Aeppli2024}, improved pulse efficiency with  
additional cooling~\cite{kwolek20,kwolek22}, and large momentum transfer to enhance the rotation response~\cite{rudolph20}.

\section*{acknowledgements}

We are grateful for contributions to the experimental setup by Jumin Lee, Isabella Campbell, Finn Woods and Eli Young, as well as helpful discussions with Joe Kramer.  
This work was supported in part by the Center for Quantum Research and Technology at the University of Oklahoma and the Avenir Foundation, an award from the W. M. Keck Foundation, and by STTR contract FA864921P0780 awarded to IFOS Corporation with the University of Oklahoma as research collaborator.  The views expressed are those of the authors and do not necessarily reflect the official policy or position of the Department of the Air Force, the Department of Defense, or the U.S. government. DISTRIBUTION A. Approved for public release: distribution unlimited. Public Affairs approval AFRL-2025-1668

\appendix*
\section{Interferometer fringe contrast}
The probability of an atom being in the ground state when it reaches the detection laser can be modeled using the Lindblad master equation.  
The system is described with two levels: a ground state and an excited state that decays back to the ground state with time constant $\tau$.  
This two level system can be described using a 2$\times$2 density matrix $\rho$; for the present calculation, the ground state is given first.
The interferometer sequence from the first beam to the detection laser is decomposed into six steps:
\begin{enumerate}
\item $\theta = \pi/2$ pulse with phase $\phi = \phi_1$
\item Decay for time $t = T$ between beams 1 and 2
\item $\theta = \pi$ pulse with phase $\phi = \phi_2$
\item Decay for time $t = T$ between beams 2 and 3
\item $\theta = \pi/2$ pulse with phase $\phi = \phi_3$
\item Decay for time $t = T'-T$ between beam 3 and detection laser
\end{enumerate}
The pulses in the odd steps can be described using a unitary transformation of the density matrix
\begin{equation}
\rho^{(n+1)} = U(\theta,\phi)\rho^{(n)}U(\theta,\phi)^\dagger\rm{,}
\end{equation}
where
\begin{equation}
U(\theta,\phi) = \left(
\begin{matrix}
\cos\bigl(\frac{\theta}{2}\bigr)& -e^{-i \phi}\sin\bigl(\frac{\theta}{2}\bigr)\\
e^{i \phi} \sin\bigl(\frac{\theta}{2}\bigr)& \cos\bigl(\frac{\theta}{2}\bigr)
\end{matrix} \right)\rm{,}
\end{equation}
while the decays in the even steps can be modeled using
\begin{equation}
\rho^{(n+1)} = \left(
\begin{matrix}
\rho^{(n)}_{11}+(1-e^{-t/\tau})\rho^{(n)}_{22}& e^{-t/(2\tau)}\rho^{(n)}_{12}\\
e^{-t/(2\tau)}\rho^{(n)}_{21}& e^{-t/\tau}\rho^{(n)}_{22}
\end{matrix} \right)\rm{.}
\end{equation}
Starting with an initial density matrix $\rho^{(0)}$ given by
\begin{equation}
\rho^{(0)} = \left(
\begin{matrix}
1& 0\\
0& 0
\end{matrix} \right)\rm{,}
\end{equation}
the steps above can be applied in sequence to compute $\rho^{(6)}$.
The upper left element of $\rho^{(6)}$, which corresponds to the population in the ground state at the detection laser, is given by
\begin{equation}
\rho^{(6)}_{11} = \frac{1}{2}e^{-T'/\tau}\cos{\left(\phi_1-2\phi_2+\phi_3\right)} + \Bigl(1-\frac{1}{2}e^{-(T'-T)/\tau}\Bigr)\rm{.}
\label{eqn:rho611}
\end{equation}
It could be argued that, due to photon recoils, the excited state does not truly decay back to the original ground state but to a new ground state with a different momentum.  
To account for this possibility, the calculation was repeated assuming decays to a new ground state in each even step.
Each of these new ground states, except for the last, was assigned a new excited state for a total of seven states.  
The total population in the four ground states after the final step ($\rho^{(6)}_{11} + \rho^{(6)}_{33} + \rho^{(6)}_{55} + \rho^{(6)}_{77}$) was found to be exactly the same as $\rho^{(6)}_{11}$ in equation \ref{eqn:rho611}.

\bibliography{main}

\end{document}